\documentclass[12pt]{article}

\usepackage{amsfonts, amsmath}

\newcommand{\be}{\begin{equation}}
\newcommand{\ee}{\end{equation}}

\newcommand{\vect}[1]{\ensuremath{\boldsymbol{#1}}}

\def\ndd{\noindent}

\begin{document}

\title{On  Geometric Realization of Quantum Computations in Externally
Driven 4-Level System}
\author{A.E.Shalyt-Margolin \hspace{1.5mm}$^1$\thanks
{Phone (+375) 172 883438; e-mail alexm@hep.by}, V.I.Strazhev$^1$
\hspace{1mm} and \hspace{1mm} A.Ya.Tregubovich$^2$
\thanks{Phone (+375) 172 841559; e-mail a.tregub@open.by}
}
\date{}
\maketitle
 \vspace{-25pt}
{\footnotesize\noindent $^1$National Centre of High Energy and
Particle Physics,Bogdanovich
 Str.153, Minsk 220040, Belarus\\
$^2$Institute of Physics National Academy of Sciences
                          Skoryna av.68, Minsk 220072, Belarus}\\

{\bf\small\noindent Abstract}\\
{\footnotesize
The possibility of realization of quantum gates by means of the non-adiabatic geometric
phase is considered. It is shown that the non-adiabatic phase can be used for
quantum gates realization as well as the adiabatic one.
}
\vspace{0.5cm}
{\ttfamily{\footnotesize  geometric phase}}

\rm\normalsize
\vspace{0.5cm}

Intensive investigations on quantum information theory (see \cite{cabello}
 for a reference source on this subject) refreshed some interest on Berry phase
 \cite{berry}. The idea of using unitary transformations produced by
the  Berry phase  as quantum computations is proposed in
\cite{zanardi} and realized in \cite{pachos1,pachos2} in a
concrete model of holonomic quantum computer. Calculation aspects
of this model are considered in \cite{margolin}. For other
references where Abelian Berry phase is considered in the context
of quantum computer see e.g. \cite{ekert} - \cite{averin}.
  On the other hand non-adiabatical Berry phase can exist and be
measured  if transitions in a given statistical ensemble do not
lead to loose of coherence \cite{appelt1,appelt2}. Therefore it is
also possible to use the corresponding unitary operators to
perform quantum computations. This fact has been noticed in
\cite{keiji}. In this paper we show a realization of quantum gates
for a concrete 4-level quantum system driven by external magnetic
field.

\ndd Let us consider  a system of two non-interacting qubits in a
bosonic environment described by the Hamiltonian
\begin{equation}\label{ham1}
 H = H_S + H_B + H_{SB},
\end{equation}
where $H_S$ is the Hamiltonian of two coupled spins
\begin{equation}\label{hamspin}
  H_S = H_S^{(0)} + H_S^{{\rm int}} = {\omega_{01}\over 2}\,\sigma_{z1}\otimes 1_2 +
        {\omega_{02}\over 2}\,1_2\otimes\sigma_{z2} +
        {J\over 4}\, \sigma_{z1}\otimes\sigma_{z2}  ,
\end{equation}
where $J$ is the coupling constant, $H_B$ is the Hamiltonian of the bosonic
 enviroment
\begin{equation}\label{hambos}
 H_B = \sum\limits_k\,\omega_{bk}(\hat{b}_k^+\hat{b}_k+1/2),
\end{equation}
and $H_{SB}$ is the Hamiltonian of the spin- enviroment interaction.
\begin{eqnarray}
  H_{SB} & = & H_{SB}^{(1)} + H_{SB}^{(2)},  \label{hamint} \\
        &        &     \nonumber \\
  H_{SB}^{(a)} & = & S_z^{(a)}\sum\limits_k\,( g_{ak}\hat{b}_k^+ +
                    g_{ak}^*\hat{b}_k) \quad a=1,2 . \label{hamsba}
\end{eqnarray}
Here
$$
   S_z^{(1)} = \sigma_{z1}\otimes 1_2,\quad  S_z^{(2)} = 1_2\otimes\sigma_{z2},
$$
$\sigma_z$ is the third Pauli matrix, $1_2$ is $2\times 2$ unit matrix,
$\hat{b}_k^+, \hat{b}_k$ are bosonic creation and annihilation operators and
$g_{ak}$ are complex constants. We assume that the two spins under consideration
are not identical so that $\omega_{01}\neq\omega_{02}$. The Hamiltonian
determined by (\ref{ham1}) -- (\ref{hamsba}) is a natural generalization of
Caldeira-Legett Hamiltonian \cite{caldeira} for the case of two non-interacting
spins.
Let such a system be placed in the magnetic field affecting the spins but not the phonon
modes. The only change to be made in the spin part (\ref{hamspin}) is the substitution
$$
     \omega_s \sigma_z \longrightarrow \vect{B}\vect{\sigma},
$$
Three components
of $\vect{B}$ represent a control set for the qubits under consideration.
Evolution of $\vect{B(t)}$ generates evolution of the reduced density matrix
$\rho_s(t)$ that describes the spin dynamics
\begin{equation}\label{rhos}
  i\,{\partial \rho_s(t)\over\partial t} = H_S\,\rho_s(t),
  \quad \rho_s(t) = U(t)\rho(0)U^+(t).
\end{equation}
Thus given curve in the control space corresponds to a quantum
computation in which each qubit is to be processed independently.
To obtain such a calculation as a function of control parameters
we first recall some common issues of spin dynamics. We consider
the external magnetic field as a superposition of a constant
component and a circular polarized wave:
\begin{equation}\label{magfield}
  \vect{B} = \vect{B}_0 + \vect{B}_1 e^{i\omega_R t},
\end{equation}
where $\vect{B}_0$ is perpendicular to $\vect{B}_1$. It is well known that the
case of the circular polarization is exactly solvable. The evolution of an
individual spin corresponding to the Hamiltonian
\begin{equation}\label{spinham}
  H = - \vect{\mu}\vect{B}
\end{equation}
 is determined by the following operator:
\begin{equation}\label{evolspin1}
 V(t) = \exp\left(\xi (t) S_+ - \xi^*(t)S_-\right)\exp ( - i\phi (t) S_z),
\end{equation}
where $ S_{\pm} = S_x \pm iS_y $ and
\begin{eqnarray}
  \xi (t) & = & |\xi (t)|\exp \left(i\Delta\omega t + i\alpha (t) + i\pi/2\right),
                           \label{xi1}  \\
                    &  &   \nonumber     \\
 |\xi (t)| & = & \frac{\omega_{\bot}\sin (\Omega t/2)}{\sqrt{(\Delta\omega )^2 +
 \omega_{\bot}^2}},         \nonumber     \\
                    &  &    \nonumber     \\
\alpha (t) & = & \arctan \left( {\Delta\omega\over\Omega}\tan (\Omega t/2)\right),
                             \nonumber     \\
                    &  &      \nonumber     \\
\phi (t)  & = &  \omega_{\bot}\,(\xi_1 n_2 + \xi_2 n_1), \label{phi1}
\end{eqnarray}
where $\Delta\omega = \omega_{\parallel} - \omega_R$,
$\Omega^2 = (\Delta\omega)^2 + \omega_{\bot}^2$, $\omega_{\bot}$ and
$\omega_{\parallel}$ are Rabi frequencies corresponding to $\vect{B}_0$ and $\vect{B}_1$
respectively and finally $\vect{n}$ is the unit vector along $\vect{B}_1$.

It is known \cite{appelt} that the pure states acquire within the rotating wave
approximation a phase factor that after one complete cycle $T = 2\pi/\omega_R$ is:
\begin{equation}\label{totphase}
  |m(T)> = \exp (- i\phi_D + i\gamma )\,|m(0)>,
\end{equation}
where $m$ is the azimutal quantum number and the phase is split in two parts:
dynamic
$$
\phi_D = 2\pi m\,{\Omega\over\omega_R}\,\cos (\theta - \theta^*)
$$
and geometrical

\begin{equation}\label{nonadphase}
  \gamma = - 2\pi m\cos\theta^*,
\end{equation}
where $\cos\theta = B_0/B$ ($\vect{B} = \vect{B}_0 + \vect{B}_1$) and $\theta^*$
is determined by the formula
\begin{equation}\label{thstar}
 \tan\theta^* = \frac{\sin\theta}{\cos\theta + \omega_R/\Omega}.
\end{equation}
The phase shift between the states $|\pm 1/2 >$ results then in
\begin{equation}\label{solangle}
  \Delta\phi_g = - 2\pi\cos\theta^*
\end{equation}
that is nothing but the solid angle enclosed by the closed curve
$\vect{B}(0) = \vect{B}(T)$ on the Bloch sphere. If the rotation is slow such
that $\omega_R/\Omega \rightarrow  0$ then $\theta^* \rightarrow \theta$ and
phase shift(\ref{solangle}) coincides with the one that is due to Berry phase
\cite{berry} widely discussed in the literature in the last 15 years
\cite{wilczek-shapere}.

Thus the adiabaticity condition is not really necessary for obtaining of the
geometrical phase in an ensemble of spins if the decoherence
time is much greater than $T$. Therefore one can attempt to use this phase to
get quantum gates such as CNOT. Calculation of the corresponding phase factors
is rather straightforward because the free and the coupling parts of the spin
Hamiltonian commute with each other
$$
\left[H_S^{(0)}, H_S^{{\rm int}}\right] = 0.
$$
Therefore the coupling part can be diagonalized simultaneously with the free
part by applying of the transformation $U = U_1\otimes U_2$ where $U_{1,2}$ are
the diagonalizing matrices for each single-spin Hamiltonian respectively. This
simple fact together with the following obvious identity
$$
  U^{\dagger}\dot{U} = U^{\dagger}_1\dot{U}_1\otimes 1_2 +
  1_2\otimes U^{\dagger}_2\dot{U}_2
$$
the final formula for the part of the evolution operator that stands for the
non-adiabatic geometric phase
\begin{equation}\label{gate1}
  U_g = \exp (-2\pi i \cos\theta_1^*\, S_{1z})\otimes
  \exp (-2\pi i \cos\theta_2^*\, S_{2z}),
\end{equation}
where
$$
   \tan\theta_1^* = \frac{\sin\theta_1}{\cos\theta_1 + \omega_R/\Omega_1},
    \quad
    \tan\theta_2^*  = \frac{\sin\theta_2}{\cos\theta_2 + \omega_R/\Omega_2}
$$
  and
\begin{eqnarray}
   \cos\theta_1 = \omega_{01}/ \Omega_1, \quad &
   \Omega_1^2  =  \omega_{01}^2 + \omega_1^2,  \nonumber  \\
                &  \nonumber  \\
   \cos\theta_2  =  \omega_{02}/ \Omega_2, \quad  &
   \Omega_2^2  =  \omega_{02}^2 + \omega_1^2 . \nonumber
 \end{eqnarray}
Note that gate (\ref{gate1}) is symmetric with respect to the spin transposition
as it should be and does not depend on $J$ that is typical for geometrical phase
in spin systems where the phase depends only on the position drawn by the vector
\vect{B} on the Bloch sphere. As $J$ does not affect this position, it is
absent in the final result. We do not consider here the dynamic phase determining
by the factor
$$
U_d = \exp \left(-{i\over \hbar}\, \hat{H}_ST\right).
$$
It is so because one can eliminate it by making use of the net effect of the
compound transformation proposed in \cite{ekert}. After this transformation
that is generated by two different specifically chosen contours the dynamic
phase acquired by the different spin states becomes the same and the geometric
phase of each state is counted twice. After that we get (up to a global phase)
the following quantum gate
\begin{equation}\label{gate2}
  U_g = \begin{pmatrix}
   e^{i(\gamma_1 + \gamma_2)}  &  0  &  0  &  0  \\
           0                  &e^{i(\gamma_1 - \gamma_2)} &  0  &  0  \\
                         0   &  0  & e^{i(-\gamma_1 + \gamma_2)} &  0  \\
                         0   &  0  &  0  & e^{-i(\gamma_1 + \gamma_2)}  \\
                           \end{pmatrix}.
\end{equation}
Thus we have constructed the quantum gate, which possess the advantage to be fault
tolerant with respect to some kinds of errors such as the error of the amplitude control
of \vect{B}. On the other hand this approach makes it possible to get rid of the
adiabaticity condition that strongly restricts the applicability of the gate.
Instead of this condition one needs some more weak one: $\tau \gg \omega_R^{-1}$,
where $\tau$ is the decoherence time.


\begin{thebibliography}{99}
\bibitem{cabello} Cabello A. Bibliographical guide to the
foundations of quantum mechanics and quantum information,
quant-ph/0012089,204P.
\bibitem{berry}   Berry M.V.// Proc.Roy.Soc.London.1984.V.A392.P.45.
\bibitem{zanardi} Zanardi P.,Rasetti M.//Phys.Lett.1999.V.A264.P.94.
\bibitem{pachos1} Pachos J.,Zanardi P. and Rasetti M.//Phys.Rev.2000.V.A61.P.1.
\bibitem{pachos2} Pachos J.,Chountasis S.//Phys.Rev.2000.V.A62.P.2318.
\bibitem{margolin} Margolin A.,Strazhev V. and Tregubovich A.
Geometric phases and quantum computations,quant-ph/0102030,6P.
\bibitem{ekert}   Ekert A. et al, Geometric quantum computation,
 quant-ph/0004015,15P.
\bibitem{pel} Pellizzari T. et al.// Phys. Rev. Lett.1995.V.75.P.3788.
\bibitem{averin}  Averin D.V.//Solid State Commun.1998.V.105.P.659.
\bibitem{appelt1}  Appelt S., W\"ackerle G., and Mehring M.//
                  Phys. Rev. Lett.1994.V.72.P.3921.
\bibitem{appelt2} Appelt S., W\"ackerle G., and Mehring M.//
                  Phys. Lett.1995.V.A204.P.210.
\bibitem{keiji}   Wang Xiang-Bin,Kieji M.,Non-adiabatic
conditional geometric phase shift with NMR, quant-ph/0101038,6P.
\bibitem{caldeira} Caldeira A.O.,Legett A.J.//Phys. Rev.Lett.1981.V.46.P.211.
\bibitem{wilczek-shapere}Shapere A. and Wilczek F. (eds.)
Geometric Pheses in Physics, World Scientific,
                               Singapore, 1989.456 P.

\end{thebibliography}
\end{document}